# Expert Elicitation for Reliable System Design[1]

**Tim Bedford, John Quigley and Lesley Walls**

*Abstract.* This paper reviews the role of expert judgement to support reliability assessments within the systems engineering design process. Generic design processes are described to give the context and a discussion is given about the nature of the reliability assessments required in the different systems engineering phases. It is argued that, as far as meeting reliability requirements is concerned, the whole design process is more akin to a statistical control process than to a straightforward statistical problem of assessing an unknown distribution. This leads to features of the expert judgement problem in the design context which are substantially different from those seen, for example, in risk assessment. In particular, the role of experts in problem structuring and in developing failure mitigation options is much more prominent, and there is a need to take into account the reliability potential for future mitigation measures downstream in the system life cycle. An overview is given of the stakeholders typically involved in large scale systems engineering design projects, and this is used to argue the need for methods that expose potential judgemental biases in order to generate analyses that can be said to provide rational consensus about uncertainties. Finally, a number of key points are developed with the aim of moving toward a framework that provides a holistic method for tracking reliability assessment through the design process.

## 1. INTRODUCTION

Statistics is considered one of the major contributors to the development of reliability engineering as a technical discipline [131]. Recent reviews of the role of statistics within reliability engineering [12, 82, 92, 102] underline the continued need for statistical science to help engineers assess sources of uncertainty, design sound data collection systems, and develop models for combining data and quantifying uncertainty. However it is also recognized that the role of statistical science within the engineering process needs to broaden to accommodate the additional complexities of the technological systems as well as the operational contexts. One particular challenge is the need to structure and integrate statistical modeling within the systems engineering process to support decision-making aimed at obtaining a sufficient and cost effective state of knowledge about future system reliability. This implies that judgemental, as well as objective, data should be collected responsibly and used formally.

*Tim Bedford, John Quigley and Lesley Walls are Professors, Department of Management Science, Strathclyde University, Glasgow G1 1XP, Scotland, United Kingdom e-mail:*
*tim.bedford@strath.ac.uk; j.quigley@strath.ac.uk, lesley.walls@strath.ac.uk.*

This paper aims to survey and review the use of subjective expert judgement methods to assess reliability in the design process. We have deliberately chosen to interpret the scope of these terms in a relatively broad fashion. Thus "expert judgement" refers to any structured method of acquiring knowledge from experts; "reliability" covers the broader issues of reliability, availability and maintainability (RAM); and the "design process" is considered to include within its scope a consideration of how the system is to be manufactured, how users will interact with it and how it will be maintained. More specifically, since reliability measures are usually expressed in probabilistic terms, we consider the use of expert judgement to structure probabilistic models and to quantify uncertainties in the development of a reliable design.

The standard definition of reliability, "the ability of a system to perform a required function under stated conditions for a stated period of time" [70], naturally translates into a probability measure. While empirical reliability can only be properly assessed after a system is in use, there is a need to forecast reliability during the design process to support analysis aimed at improving reliability. Davis [33] supported the definition found in [23] that "reliability is failure mode avoidance." We are sympathetic to this view since identifying and mitigating influential critical failure modes will cause reliability to improve. However, we also believe that probabilistic models have an important role to play in supporting design decisions since they allow data integration and assist prioritization.

Reliability is a recognized element of systems engineering and systems design. However, it is worth recognizing from the outset how difficult it is to talk about the reliability of a system. In part the difficulty has to do with ambiguity of any reliability metric. In modern systems engineering the practice of requirements setting should, if carried out well, result in a coherent set of reliability requirements expressed in terms of well-defined RAM metrics. Hence good engineering-management practice should ensure that there is little ambiguity in the expression of reliability requirements. More difficult though is the uncertainty around the circumstances under which those requirements are to be met. The reliability of a system is ultimately determined by a combination of factors. Simplistically, we may think of the reliability of a specific system as being determined by the detailed design reliability as modulated by induced unreliabilities coming from the manufacturing process, from the users, from maintenance and from modifications. Simplistically, detailed design reliability gives the maximum potential reliability which manufacturing errors, poor usage and poor maintenance will typically reduce, while changes or modifications introduced as a result of experience with the equipment will improve the reliability, that is

overall reliability = designed reliability

$$- \text{ production unreliability}$$
$$- \text{ usage unreliability}$$
$$- \text{ maintenance unreliability}$$
$$+ \text{ changes reliability.}$$

More compactly, we could write a chosen reliability measure $r$ as

$$r = r(d, p, u, m, c),$$

where $d, p, u, m$ and $c$ represent the choices made for design, production, usage, maintenance and changes. Inasmuch as systems engineering is about making trade-offs between different aspects of the system, the major focus for expert judgement techniques in support of reliability has to be to explore the behavior of, and even quantify, the above conceptual function in some way.

The existing expert judgement literature is a starting point for elicitation problems in engineering design, but it needs to be extended to cope with the unique problems encountered. This is one of the motivations for the present paper. In discussing the ways in which expert judgement methods are adopted to assess uncertainties in the design process we shall consider both the academic and foundational aspects as well as the typical business context so that we can gain an understanding of why simpler methods are not replaced in practice by better founded methods.

The paper is structured as follows. After describing the systems engineering life cycle phases, we examine the role of the stakeholders within key markets and their influence on reliability modeling intentions. Summaries of the existing literature in elicitation are woven into a discussion of the issues that arise during model structuring, instantiation and updating across the systems engineering process. We conclude by suggesting areas in need of further research.

## 2. SYSTEMS ENGINEERING AND DESIGN PHASES

Systems engineering is described in the *NASA Systems Engineering Handbook* [132] as

> ...a robust approach to the design, creation, and operation of systems. In simple terms, the approach consists of identification and quantification of system goals, creation of alternative system design concepts, performance of design trades, selection and implementation of the best design, verification that the design is properly built and integrated, and post-implementation assessment of how well the system meets (or met) the goals.

Reliability is regarded as an important specialism that supplies expertise to the systems engineering process [14, 15]. However, the nature of reliability knowledge and the demands placed on its practitioners changes considerably through the systems engineering process. It is therefore useful to consider the main stages in the design process.

### 2.1 Life Cycle Phases

The phases described here, based on the most recent international standards [69], are generic, but descriptions of design phases vary in the literature [14].

- *Concept and definition.* Requirements definition is the generation of technical design constraints on the system. Some of these will be derived from information about user demands or expected user wishes, while others will be there to ensure feasibility of the design. Trade-off studies are carried out in order to achieve cost-effectiveness and feasibility. Finally, initial life cycle costing studies will be made.
- *Design and development.* The system architecture is specified in detail, hard- and software will be built, tested and refined, leading where necessary to adjustments of the specification. Verification and validation of subsystem integration is carried out: verification ensures that subsystems interfaces conform to design specifications and validation ensures that the integrated systems fulfill their intended function. Maintainability analysis will be carried out and end of life disposal will be considered.
- *Manufacturing and installation.* Hardware will be produced and software will be replicated. There is an emphasis on process control, although further product verification and validation will take place. Field trials may be used as a final check on system performance.
- *Operation and maintenance.* The system in use should be monitored for performance. Maintenance also provides clues as to system performance and can be adjusted where necessary.
- *Disposal.* Depending on the regulatory context, the system may be destroyed, dumped or dismantled. Increasingly there is pressure from regulatory authorities for reuse of equipment subsystems, so this stage is by no means the end for the system components.

Figure 1 illustrates the relationships between the different systems engineering phases. Prior to operation, reliability estimates forecast true performance and will encompass the uncertainties in future decisions. As system-specific observations are collated during development and manufacture, some uncertainties should be resolved and reflected in revised estimates. This is shown schematically in Figure 2.

Feedback loops exist both within and between phases, reflecting the analogy with a control system. Figures 3–6 show more detailed activities with each flow chart capturing the cyclic nature of the process to refine the system design based on assessment against reliability requirements. Information from subsequent phases should be fed back to earlier phases with a view to modifying the current design, if required, but also to inform processes and procedures that will impact later generations. However, feeding data backward is only possible when the systems engineering phases overlap. Hence much of the data being fed back is judgemental in nature. We shall return to these flow charts later when we discuss issues relating to the role of elicitation.

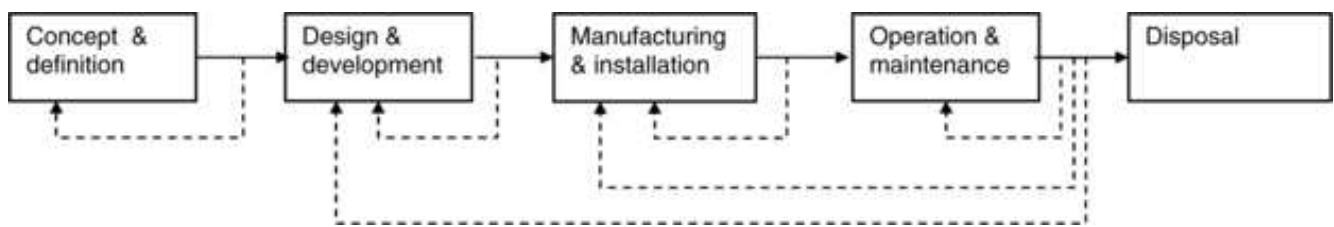

FIG. 1. *Systems engineering phases.*

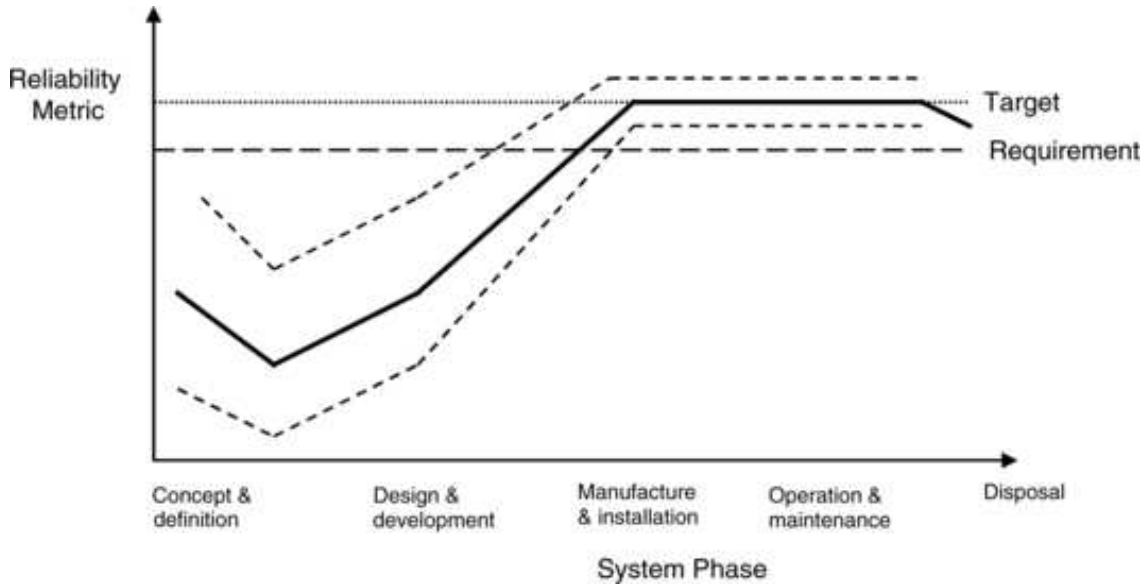

Fig. 2. *Decreasing uncertainty in future system reliability.*

### 2.2 Stakeholders in System Design

As mentioned above, these phases are generic and hence relevant to the markets for consumer, industrial and military systems. However, there may be differences in the nature of reliability knowledge and modeling within each market, and we explore this further through consideration of the key stakeholders with interests in following the reliability assessment of a new system, namely requirements specification team, design team, component manufacturer, lead manufacturer, sellers, regulators, end users, general public, maintainers and disposers/recyclers.

These parties can be classed within one of four groups: client, manufacturer, regulator and public. These stakeholders can, and often do, take different viewpoints about the reliability of the system and about the relevance of data. For example, Table 1 captures the respective roles of the groups and aims to illustrate two key points. First, that different stakeholders may have different modeling intentions with, for example, manufacturers using models to measure reliability to support decisions about accommodation of failure modes and improvement activities, while clients may use models to support negotiation with manufacturers. Second, that during such negotiations different stakeholders may be using the same data to support different sides of a decision. This latter situation mirrors a similar situation in probabilistic risk assessment and indeed areas where different parties are asked to adopt a common view of uncertainties. This was Cooke's motivation for the notion of "rational consensus" [27]. See also the extensive literature on risk communication and public perceptions, for example, [54, 127, 128, 133, 136].

## 3. ELICITATION IN RISK AND RELIABILITY

Subjective expert judgement has a very important role to play in assessing uncertainties in the design process, with many of the stakeholders identified above contributing their expertise. However, the emphasis is somewhat different from the role that expert judgement has in other areas—most notably in probabilistic risk assessment (PRA). Much of the modern academic literature on expert judgement has emerged from the need for structured subjective assessments in PRA. The key issues that emerge from this literature are reviewed.

### 3.1 Roles within Elicitation

In principle there are three distinct roles:

- *Decision-maker*: This person is the problem owner, who is responsible for signing off on a decision and

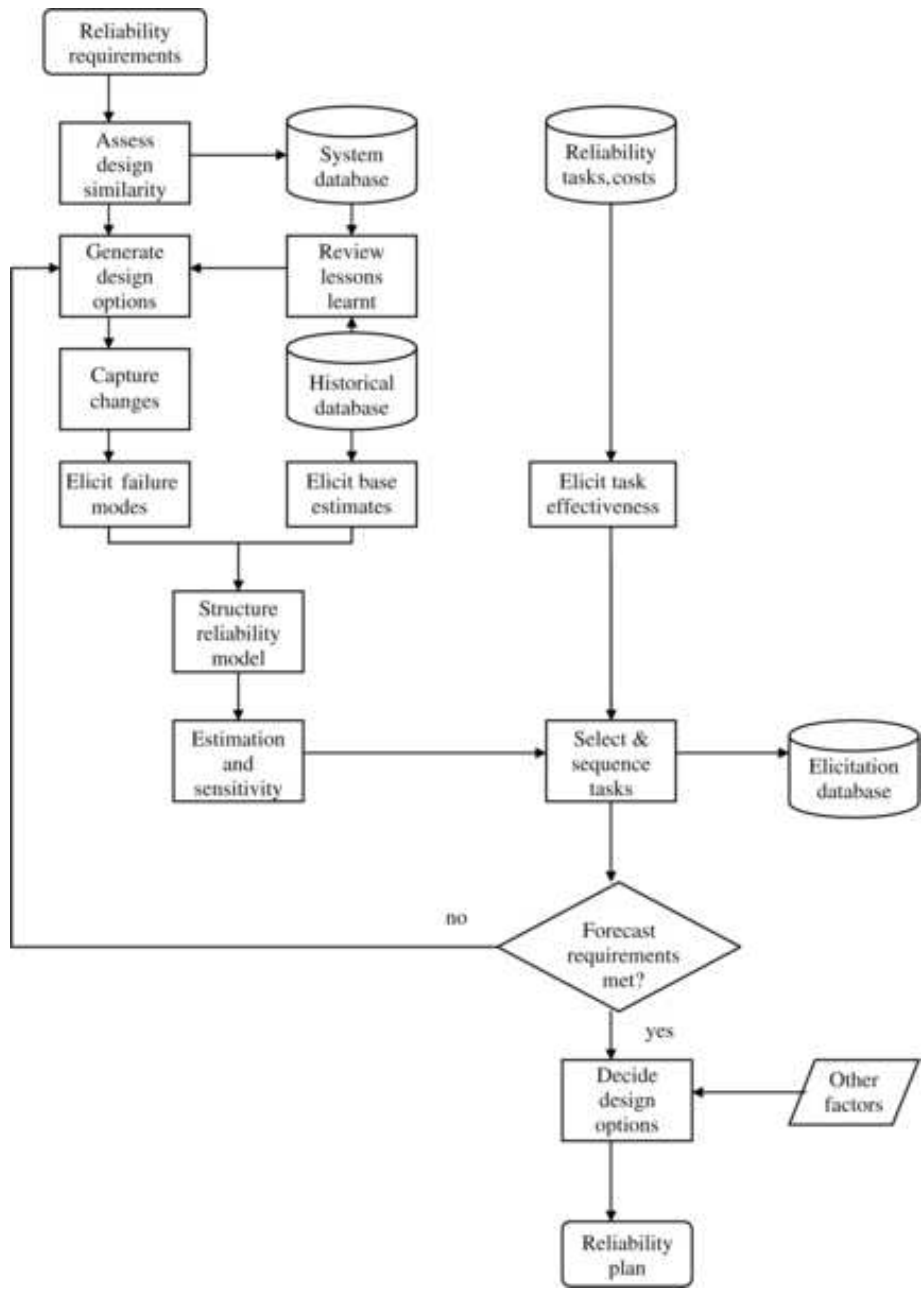

Fig. 3. *Concept and definition flow chart.*

wishes to be informed about relevant uncertainties by appropriate experts.

- *Expert*: This person is identified as a domain expert and contributes his or her own assessment on the events of interest.
- *Analyst*: This person is responsible for identifying experts and events of interest, and writing the assessment and combination schemes.

It should be noted that [17] also distinguishes the role of *advisor-expert* who essentially plays a role somewhere between the above players, by supporting, for example, the selection of experts and elicitation questions. The distinction between roles is valuable because some schemes do not recognize the different roles of these players. Bayesian schemes in particular often merge the role of expert and analyst by requiring that the analyst play the role of meta expert, by providing a prior that the expert data will subsequently update and/or by providing the likelihood function for the expert data.

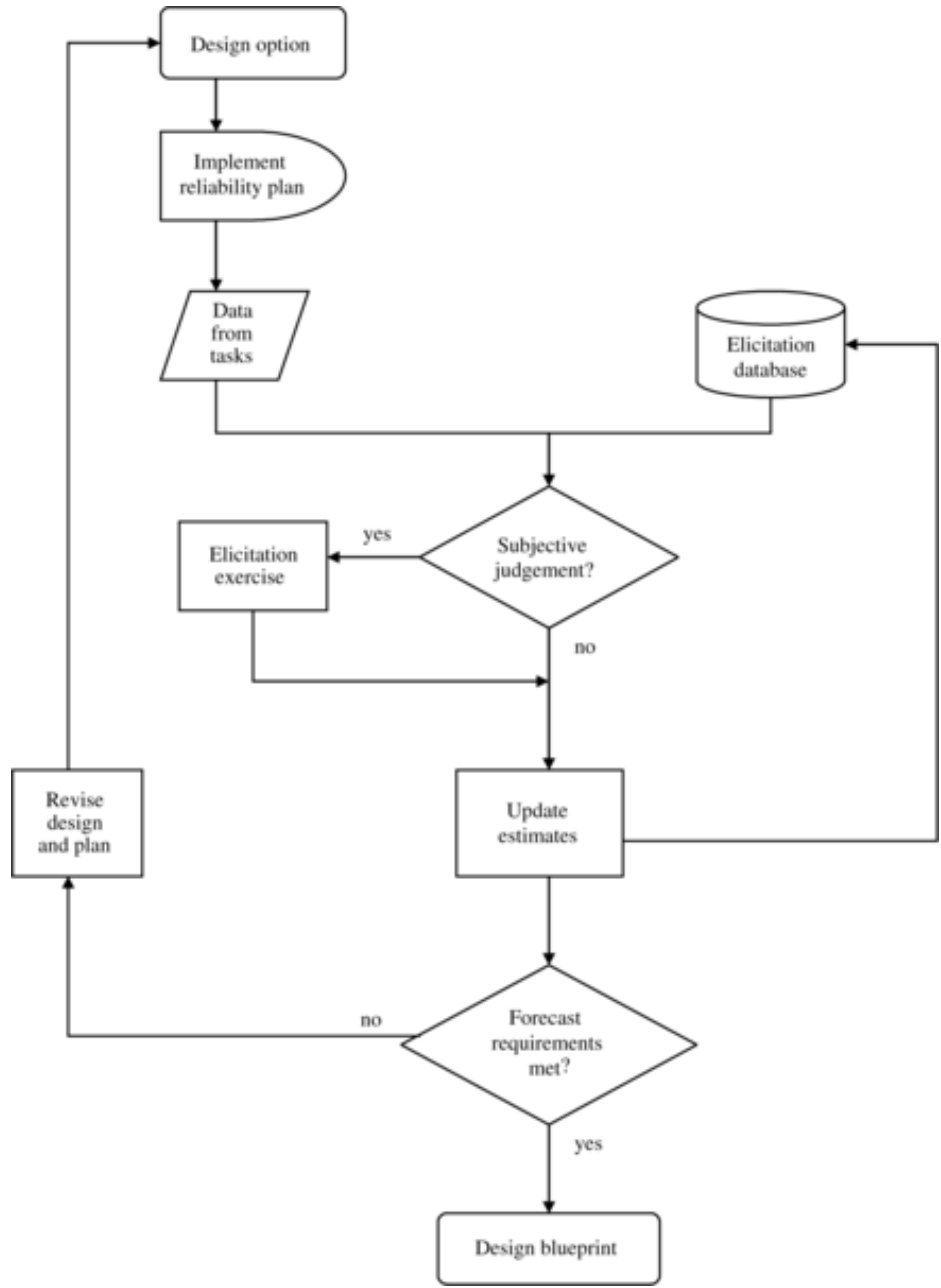

FIG. 4.   *Detailed design and development flow chart.*

A distinction between the three roles defined above would seem to be important, for example, in public sector decision-making where there is a need for transparency. Even in the private sector there is a benefit to be gained from transparency and a clear division of roles. However, it clearly also imposes a cost burden, for example, due to the degree of specialism involved, and may therefore be less appropriate in some contexts.

### 3.2 Probability Elicitation Methods and Processes

Research in experimental psychology has demonstrated that accurate subjective probabilities are unobtainable by simply asking someone to provide a probability number; therefore an elicitation process is required [75, 103, 104]. Much of the research in elicitation is concerned with minimizing bias, which can result from a variety of causes. Four standard forms of bias are: *motivational*, which concerns the situation where the expert has an interest in a particular value for the parameter being assessed; *cognitive*, which can result from incoherently basing an assessment on a number of calculations; *anchoring*, which exists when assessments are derived by an expert from adjusting previous assessments; and *availability*, which concerns assigning higher likelihoods to events that are linked to more memorable historical events.

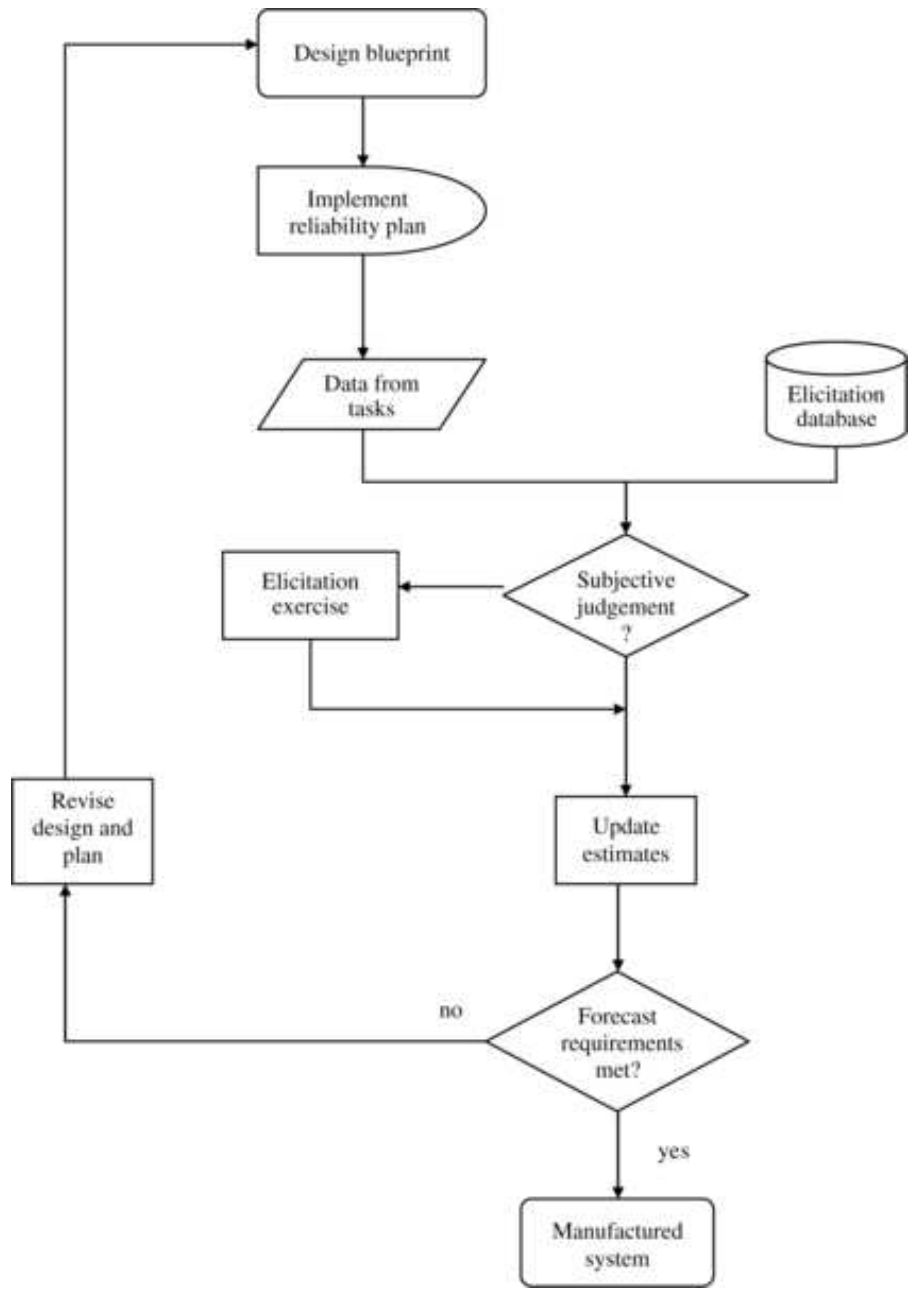

Fig. 5. *Manufacturing and installation flow chart.*

Clemen and Winkler [24] gave an overview of the state of the art with particular emphasis on risk analysis applications. O'Hagan and co-workers wrote a series of papers which probably encompass the most recent generally applicable work on elicitation [53, 72, 113, 114, 115]. An overview of the uses of expert judgement in engineering applications was given by Ayyub [7], although this work covers also aspects of fuzzy representations (about which the reader can find a review by Cooke [29] of a previous book by Ayyub and a reply by the author). Fitting closely to the theme of this paper is the work of Booker and McNamara [17], which presented a very nice description of the process of determining problems for expert judgement, selecting experts and the problems caused by possible biases. Cooke [27] gave an historical account of elicitation and also provided a number of different models for the combination of expert probability assessments, including the classi-cal method which has been quite successful in PRA applications; see [76] for a list of applications.

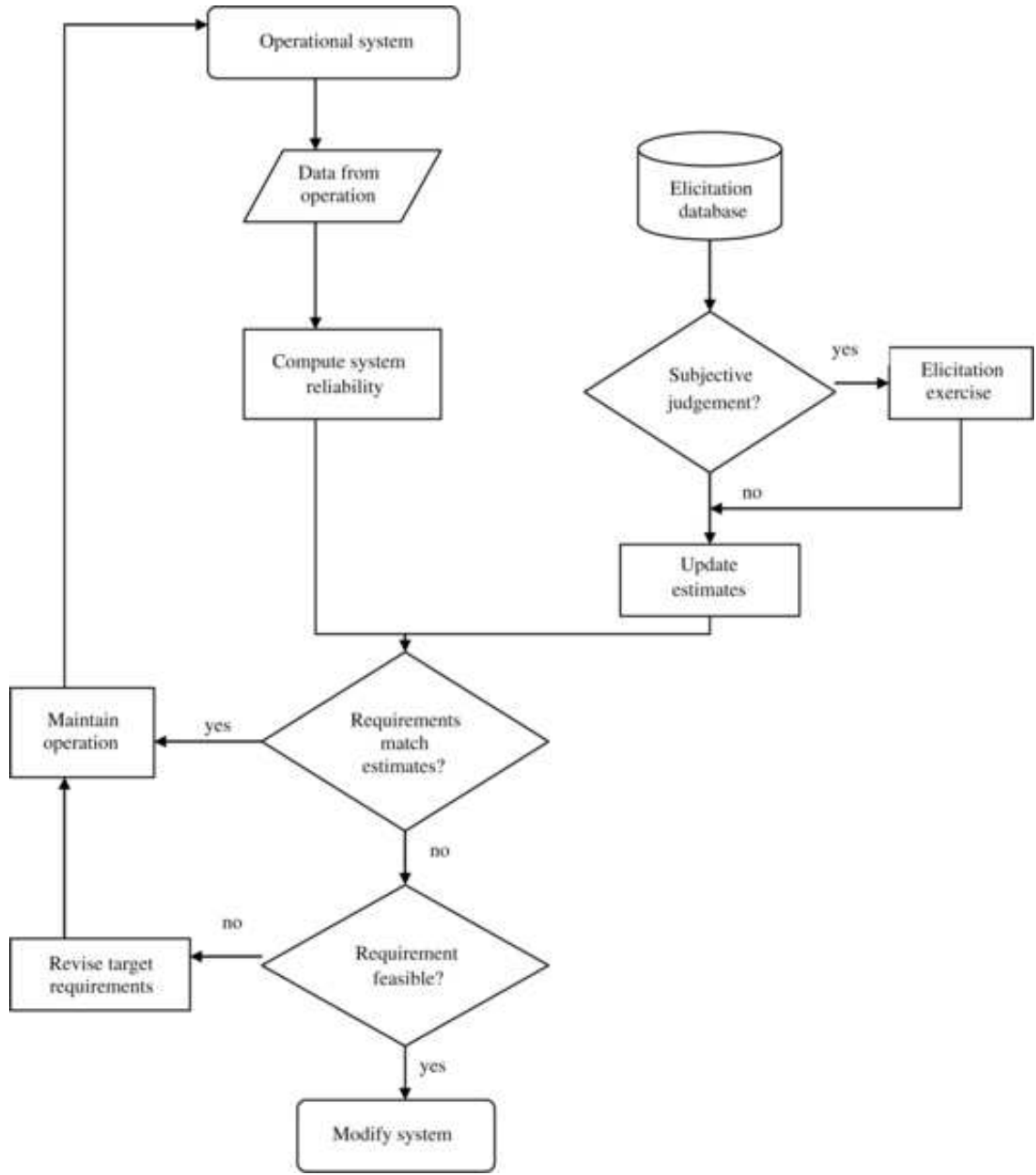

Fig. 6.   *Operation and maintenance flow chart.*

Expert judgement methods that draw on PRA are relevant to work on engineering design problems but are limited in two important ways. First, in the engineering design process there is a greater need to have experts define the problem structure, so the qualitative phase of model building is relatively more important than it has historically been for PRA decision support. See, for example, Walls and Quigley

TABLE 1
*Stakeholder uses of reliability assessment*

| Manufacturer | Client | Regulator and Public |
|---|---|---|
| Acceptance of requirements | Specification of requirements | |
| Design of reliability program | Acceptance of reliability program | |
| Proof of meeting targets | Assurance of meeting targets | Safety case |
| Planned maintenance specification | Spares ordering | |

[146], who proposed an elicitation process to support reliability growth modeling. Second, there is a big difference in the way that events can be described: PRA elicitation is generally for very precisely defined events, while in engineering design problems it is much more difficult to describe events precisely because of extra uncertainty caused by the effect of future decision-making that surrounds the system and its use. In both cases there will always be unspecified states of the world for which the expert has to "fold in" his uncertainty. However, the degree of influence of future decision-making in the reliability engineering context is such that it becomes useful to model this explicitly in order to support the design process.

These concepts are represented in the flow charts shown in Figures 3–6. For example, in the concept and definition phase we distinguish between the elicitation of the qualitative failure modes and the quantitative reliability estimates. Furthermore, sensitivity analysis represents the exploration of future uncertainties using engineering judgements as inputs. In subsequent phases, previous judgements will be revisited and revised in light of observations from analysis and test tasks.

### 3.3 Modeling Uncertainty in Design

While systems engineers may well think in the holistic framework outlined in Section 2 and captured within the flow charts in Figures 3–6, the statistical modeling generally applied is usually focussed on tightly defined and highly specific issues within life cycle phases. The support that these tools give engineers is therefore fairly constrained.

Uncertainty is fundamental to systems modeling and is worthy of further comment. Various authors have given overviews of different "types" of uncertainty. The classification given in [10] discusses aleatory and epistemic uncertainty, and suggests that the important distinction between them is model-dependent, as epistemic uncertainties are uncertainties that we wish to capture and adjust within a model through learning, whereas aleatory uncertainties are not adjusted within a model. The uncertainty in an abstract parameter or in a model type can be given an interpretation, according to [10], only in terms of the uncertainty it induces in observable outcomes. The above types of uncertainty can be quantified by subjective probability. By contrast, [10] mentions ambiguity (which is best resolved by careful definition during qualitative problem structuring rather than mathematical modeling) and volitional uncertainty (an individual's own uncertainty in his own actions), which cannot be measured by the tools of subjective probability, although it can be assessed by an independent observer.

In the context of engineering design it seems useful to define another kind of uncertainty—that of tolerance uncertainty. This represents the variation expected in a parameter across the design envelope. For example, one might be interested in the failure rate (assumed constant) associated with a piece of equipment. Since that failure rate will depend on various design, construction, environmental and usage factors assumed in the definition of the design envelope, we can write it as $\lambda(e)$, where $e$ represents chosen factors.

Assuming that $e$ is constrained to lie in a design envelope $E$, the tolerance uncertainty associated with $\lambda$ and $E$ is the interval

$$\left[\min_{e \in E} \lambda(e), \max_{e \in E} \lambda(e)\right].$$

It does not always make sense to place a probability distribution on $E$. This is because some variables are subject to choices made by the designer, the manufacturer, the user or the maintainer. (At the simplest level this could be a mandatory rule to the user to avoid certain conditions that are known to induce failure.)

System engineering places great emphasis on making trade-offs between different aspects of the system—cost, functionality, reliability and so forth. From a reliability point of view, one of the principal ways in which this trade is carried out is by changing the design (which may have cost and/or functionality implications), by specifying changes to the design envelope (i.e., restricting the way in which the system can be used), by specifying changes to the maintenance regime or by making changes and modifications to the system. Since many of these implementations occur after the design process has been (notionally) completed, to make the trade-off in the best way possible it is necessary to know how much the tolerance uncertainty can be controlled by changes made later.

## 4. ELICITATION WITHIN (RELIABILITY) MODELING PHASES

Reliability models, as is the case with many other model classes, are developed and applied through three modeling phases. These phases can occur at any point within the systems engineering process, depending on the question at hand. The conceptual phase is one of model structuring in which a qualitative form is given to the model. That is followed by an initial quantification stage and then by a revision stage in which increasing quantities of real system data can be utilized.

In all three modeling phases there is a role for expert judgement. In the first, the primary role is in model selection and initial qualitative specification. In the second, expert judgement has the key task of providing the initial quantitative estimates. In the final phase, expert judgement plays an important role in interpreting the relevance of available data.

We discuss below the roles that expert judgement plays in these modeling phases, but first we discuss frameworks described in the literature that aim to stretch across both modeling and systems engineering phases.

### 4.1 Meta Modeling Frameworks

The programs PREDICT (Performance and Reliability Evaluation with Diverse Information Combination and Tracking) [83, 84] and REMM (Reliability Enhancement Methodology and Modelling) [148] are two modeling frameworks used to estimate reliability throughout the systems engineering phases. Both models begin with a problem structuring stage, which consists of eliciting a graphical representation of the relationships between relevant engineering concerns or potential failure modes and the reliability experienced by the system. These graphs form the structure of the stochastic model; this essentially represents a meta model within which standard probability models can be integrated. The stochastic model is populated with either expert judgement or relevant historical data. Thus these approaches provide one unified decision support framework throughout the system design and development, supporting sensitivity analysis as well as credible intervals of the uncertainty in the reliability. Furthermore, as system-specific data become available through analysis and test, the model parameters can be updated.

Such frameworks rely on expert judgement for the reliability assessment at a system level and aim to overcome the limitations of traditional approaches, which according to [62] and [73] tend to provide overly optimistic estimates of reliability due to their failure to account for major sources of early failures such as design defects, process flaws and human error.

We move now to a discussion of the three modeling stages.

### 4.2 Qualitative Model Structuring

We distinguish between four types of structuring activity that play a role within the design and development phases: capturing and defining requirements, eliciting failure modes, selecting model formulations and robust design.

4.2.1 *Requirements capture and concept definition.* Reliability requirements drive the modeling process as shown in Figure 3 because they inform targets against which reliability estimates will be compared. Reliability requirements are expressed in a fairly standard form in most engineering design projects. O'Connor [111] provided guidelines of what should and should not be included. Since reliability requirements can drive significant costs, they should be motivated and ideally derived from user demands about the system functionality and from an understanding of what the current technology levels can support. However, such a derivation requires many assumptions about the pattern of use and the environment in which that will take place.

While it is acknowledged by designers of hardware systems that the customer's requirements of the item are of paramount importance [71, 111], there are few recent published articles compared with requirements setting for software systems [108]. In our experience with hardware systems it seems that systematic modeling is not performed in the derivation of requirements, and historical precedent (i.e., the requirements that were set for the last version) is used as an alternative.

A focus of research within the software community has been the elicitation, analysis and management of system requirements. Two dominant, but complementary, methods for analysis are goal oriented [35] and use case analysis [2]. The former is concerned with eliciting system constraints, while the latter is concerned with eliciting system behavior [151]. Processes have been proposed to support creative thinking about requirements [98] and capture stakeholder views [32, 39]. Comprehensive rigorous processes for requirements definition have been suggested, for example, in [158] and [6].

A further approach of value in reliability is QFD [130, 139], which provides a broad-brush, semiquantitative assessment of the relationship between those factors that can be controlled by engineering design and those characteristics valued by users.

A study of requirements changes throughout a project is given in [97]. The problem of so-called requirements creep can be endemic, and modeling the development of requirements throughout a project is not easy. Within the context of software development Stallinger and Grünbacher [137] explored modeling this with system dynamics; see also [63, 64, 65, 66].

### 4.2.2 *Eliciting failure modes.*

Qualitative reliability modeling is routinely conducted during concept design to elicit and structure the failure modes that are likely to drive the (un)reliability. Methods used include failure mode and effects analysis (FMEA) [20], which obtain an understanding of the ways in which different types of failure can occur, while hazard analysis [85], top-level event tree (ETA) and fault tree (FTA) analyses [4] can give an indication of how the system functions. These types of analysis are prospective and can be extended in later stages when more information is known about the system. In contrast, root cause analysis [38] provides a process for retrospective forensic analysis of observed events to identify the drivers so that lessons learnt can inform use and maintenance of the operational system; however, such data can also inform design modifications to a new generation.

Elicitation of subjective judgement plays a pivotal role in such qualitative analysis with all methods using some semistructured process to gather and organize data. For example, FMEA aims to develop a model of the causes, modes and effects of failures as they impact the system through different levels of indenture. Conceptually, FMEA aims to populate an exhaustive sample space of potential events that could impact reliability from a design or process perspective. The approach to elicitation is to frame questions either in terms of functionality, architecture or process and systematically think through each level in a bottom-up (i.e., from parts to system) manner. In contrast, FTA assumes a top-down approach to elicitation. Critical events, or so-called top events, are defined in terms of departures from requirements. For any system there may be one or more top events. For each, a tree is constructed by drilling down the sequence of events that could cause or exacerbate a failure. Fault trees can accommodate failures with more than one cause, while FMEA cannot. Hazard analysis represents a structured elicitation of potential operational hazards to a system during installation, production and decommissioning using a set of prescribed keywords to manage the content analysis.

The principles of the aforementioned approaches aspire to be systematic; however, there has been criticism of their reported implementation. The FMEA has been criticized within the aerospace industry [100] because it has been implemented too late in the product development process and in a manner that does not allow information to be fed back to inform the product design. White [153] criticized the general approach to use of the suite of standard methods, claiming the manner of their use is reductionist, and proposed that a systems approach that exploits multiple partial views and explores the problem environment would result in richer information.

It is not known how valid these criticisms are for all industries. There is evidence that these methods are being used effectively to influence system design of, for example, space systems [52], but there is undoubtedly a lack of reporting in the literature by manufacturers and there is no known scientific survey of the effectiveness and efficiency of their application. Anecdotal evidence suggests that there are industry effects, for example, consumer products that embrace elicitation of failure modes as part of their quality processes [33, 119] while others largely remain accountants of failure modes.

Recent research related to these qualitative methods has been dominated by two avenues: (1) automation of knowledge capture and representation [93] and (2) quantitative prioritization rules [21] and computational algorithms [5, 43]. An exception has been the work described in [17, 62, 83, 84, 147, 148] which developed elicitation processes that embrace a systems approach and the scientific principles of structured judgement [27] fundamental to sound data collection. They aim to elicit the core concerns held by all relevant stakeholders during early design through a sequence of semistructured interviews using simple mapping [41]. Discussions are triggered by focussing on the changes between generations of systems designs in terms of technology, process and use. The maps developed capture the reasoning trail that links to formal records within a defined failure taxonomy [50] that can be revisited and updated as the design evolves. This approach is captured within Figures 3–6 through the initial activity to elicit failure modes in early design and subsequent elicitation exercises in later phases.

4.2.3 *Selecting and structuring models.* Surprisingly little has been written about the qualitative structuring process for reliability models. At a practitioner level, guidance on the selection of tools to match modeling objectives exists within international and company standards. At a more abstract level, the principles of requisite modeling can be applied [120, 121, 123].

The standard systems reliability models are all based essentially on cause and effect, and include FTA, ETA, reliability block diagrams (RBD) and (semi) Markov modeling [16, 126]. These methods provide subtly different, but related, representations of the system. Typically used in a hierarchical way, they can be used at different levels of system indenture, enabling the reliability engineer to fill in more detail, as it becomes known. Keller and Modarres [81] provided details of their early history. See also [31, 99, 143].

There is often a perception that there exist "correct" models which can be found by the application of appropriate quality control. However, there are important choices to be made about the model scope. How deep (or detailed) should the model be? What failure events should be considered? Which environments, or failure scenarios, should be considered? These questions are the subject of expert judgement, albeit usually unstructured.

Graphical based methods such as FTA and RBD are popular during design because they provide useful representations of the system, linking probabilistic assessments with physical structure and functionality. However, the frameworks are not without shortcomings and recent research has proposed the use of Bayesian belief networks (BBN) as a more flexible substitute [134]. The BBNs can be constructed to directly map onto potential engineering decisions [11]; they can be constructed to capture temporal effects [18]; they can capture common cause failure modes [140]; they can capture anticipated changes in reliability due to manufacturing and operational demands [152]; and, finally, BBNs can be used to facilitate decision-making subject to multiple criteria [47]. This is important during concept design when the strengths and weaknesses of design options are traded off. Several case studies describe the application of BBNs to reliability modeling of complex systems. See, for example, [19, 46, 107, 159]. Leishman and McNamara [94] described an ethnographic approach to qualitatively structuring a reliability model. Such an approach makes use of in-depth interviews with relevant participants. The data acquired through the interviewing processes are structured via Bayesian networks; see also [155].

4.2.4 *Robust design.* The stress–strength relationship is core to reliability engineering. Conventional modelling, as discussed above, provides estimates of whether the system design possesses the strength to meet the nominal stresses within the specified operational environment. However, there can be considerable uncertainty about the actual stresses encountered in operation and, hence, analysis to examine the robustness of the design to variation in stresses is important.

The concept of robust design [116] is fundamental to the quality movement and encompasses the work on experimental design and analysis. Condra [25], among others, defines reliability as "quality through time" and advocates the importance of statistical experimental design in reliability improvement. There are limited reports of its use in practical reliability engineering, although see, for example, [33] for its use within the automotive industry. Perhaps this is not too surprising since the ability to replicate repeated trials is most feasible for those systems which will be mass produced. Others have discounted the influence of experimental design on traditional reliability testing because of the identifiability problems given the small amount of data relative to control parameters [79]. The increasing role of simulated experiments may remove such physical constraints.

Elicitation is required to support not only design of experiments, but also specification of standard reliability tests, such as growth development tests and production acceptance tests. Again there is little reported about how this can and should be achieved. Exceptions are Condra [25] and Davis [33], who share insight into the identification of the failure modes that influence the choice of response variable and the semistructured methods used to identify the explanatory variables and their experimental settings.

Methods for elicitation of stakeholder judgements abound in the quality literature. A useful summary is given in [78], which summarizes 100 methods by purpose, when to use, how to use and benefits, and provides an example. The methods are classified into: management methods, analytical methods, idea generation, data collection, analysis and display. While the scope is comprehensive, all tools are treated as independent entities.

A recent special issue of the journal *Quality and Reliability Engineering International* (April 2005) provided some interesting reviews of the role of six sigma in the 21st century and the key interaction between the softer and harder aspects of statistical modeling within industry. Hahn [60] emphasized the key goal of designing products with long life and high reliability, and identified the need to include reliability modeling within the six sigma toolkit. The use of six sigma through the life cycle of an automated decision support system is discussed in [117], again highlighting the synergies with systems engineering, although broadening the issues beyond the engineering to include service processes. Anderson-Cook, Patterson and Hoerl [3] described graduate training with special emphasis on the role of structured problem solving within a program that aims to develop the facilitation skills of statisticians within a project life cycle.

Experiments, tests and statistical quality control are encompassed by the generic term "task" used in Figures 3–5. We propose that their value should be assessed during reliability planning and the data from their implementation should be used to revise modeling estimates, which we shall discuss further later.

### 4.3 Initial Quantification

Most of the key probabilistic models used in reliability are quantified through mixtures of expert judgement and generic, or other, surrogate data.

#### 4.3.1 *Reliability models.*
Before discussing the various techniques used for quantification, we give a brief overview of some of the models used, arranged according to the systems engineering phases. Note, however, that there is no rigid restriction of models to the phases we have associated them with, as preliminary studies are frequently carried out in earlier phases and detailed later. For example, decisions about production, maintenance and operational support will tend to be made in development using information about the failure modes elicited in design. In turn, tasks included in the reliability plan and used to revise estimates after implementation include the engineering analysis and test methods discussed below.

*Design and development.* Concept design is characterized by the need to make trade-off decisions: cost against functionality, weight against strength and so on. In principle reliability requirements should play a part in these trade-offs too, with model predictions being inputs to the game. However, although there is a wide literature on reliability optimization (see [91] for a survey), this literature generally makes the assumption that the system structure and the reliability characteristics of parts are quite well defined. This is not usually the case within early design: hence, the difficulties of predicting future reliability quantitatively are such that reliability tends not to play a major role in the trade-off discussion [14].

In addition to the systems reliability models listed in Section 4.2.3 and widely used in practice, many probability models have been reported in the literature. For example, Singpurwalla [135] provided a taxonomy of stochastic models that are useful for reliability modeling in dynamic environments. This is important because not only are there uncertainties in the operational stresses under given conditions, but there can be anticipated variation on the demand patterns. Renewal processes are commonly used [8], although other people adapt FTA to capture a dynamic environment [1].

Physical failure modeling is used extensively within simulation during detailed design and development. Mathewson et al. [101] provided a review of simulation tools used within the design process for predictive inference as well as for supporting optimal design decisions. The majority of these models make extensive use of component level physical models which are adjusted by empirical data for calibration. The main criticism of these models is limited focus of one failure mechanism per model [13].

Engineering testing remains a staple part of reliability programs, but growth testing is now more prevalent than demonstration testing. Many tests will be conducted under accelerated conditions (see [109, 110] for a bibliography of accelerated test plans) and they generate few observations. Consequently, research in this field is dominated by Bayesian approaches; see [44, 58, 77, 124, 125]. A notable exception is modeling with covariates [40].

Within civil engineering, expert judgement is now commonly used to incorporate assessments of uncertainties into design decision-making. In this area much of the design decision-making is in the context of the management and upgrading of existing assets. See [34] for a discussion of a performance-based asset management system for flood defenses which is driven by expert judgements. The Dutch are going through a process of reevaluating their risk criteria for dikes, and much of the technical preparatory work has involved the use of expert judgement to assess uncertainties in the physical models of dike failure [30, 145]. Similarly, expert judgement has been used to quantify physical models that describe the behavior of buildings [36].

*Manufacture and installation.* There are a few unique systems where active design continues into manufacturing—mainly in space systems and civil engineering structures. However, for most systems, the emphasis in the manufacturing phase is on production quality control. For mass production, established methods of statistical process control can be used for the key failure modes elicited during design and development. Systems with a low volume of assembly and many manual processes may rely on product screening [67]. For such systems, which include aerospace and military systems, early production models can also be used in prerelease testing to give either the manufacturer or the client confidence in the reliability of the product. The design and analysis of these trials possess the same data challenges as reliability demonstration tests earlier in development.

*Operation and maintenance.* Several authors have acknowledged the role of expert judgement within maintenance modeling, notably, Lu, Wang and Christer [96], who combined subjective judgement about preventive maintenance with failure records to support delay time modeling of plants, and Kunttu and Kortelainen [90], who presented a case study using expert judgement within a Poisson model to support maintenance decisions. van Noortwijk et al. [142] proposed a maintenance optimization model and used a linear pool to combine expert opinion to assess the lifetime distribution. See also [149] for a review of subjective estimation in maintenance modeling.

Murthy, Solem and Roren [106] provided a comprehensive review of warranty modeling, and Kleyner and Sandborn [86] provided a warranty model based on Weibull and exponential models where the parameters are estimated by data using stochastic simulation to overcome mathematical intractability. Ward and Christer [150] acknowledged the need for expert assessment for warranty modeling. Examples of Bayesian approaches include [68, 122, 138].

Real-time condition monitoring is an important tool in maintenance decision-making. When modeled, a degradation signal can be used to estimate residual life. The data obtained through measuring aspects of the degradation process of each of the system's components can be used as concomitant variables in a proportional hazards model [144]. Alternative approaches include [87], which uses a Markov chain to capture the degradation process. The usefulness of engineering judgement for interpreting such data is evident and, as such, Bayesian methods potentially play an important role in this part of the cycle (see, e.g., [55]).

A variety of problems are associated with civil engineering structures during the operations phase. Assessments of the times required to evacuate a dike ring are made in [9], while the time required to safely close the movable barriers in a dike ring structure is modeled in [141]. Degradation process modeling is very important, particularly where inspection or condition monitoring may be costly, such as with sewers [88], or where the underlying processes are difficult to predict, such as with coastal erosion [61].

### 4.3.2 *Expert judgement collection.*

All of the above models require instantiation. Typically they are quantified through expert judgement using a similar set of techniques that we now discuss.

The initial quantification of reliability models is, in practice, frequently an unstructured search through historical systems data and generic data bases to find "ball park" parameter estimates.

A variety of problems are encountered here:

- The combination of opinions of different experts.
- The transformation of combined data into assessments of parameters within a model.
- The combination of expert and generic system data.

4.3.3 *Expert combination.* When more than one expert is contributing assessments about a quantity of interest, then the analyst has the problem of combining them in some way. There are, broadly, two approaches to this. The first is to pool the data—generally through the use of a linear pool. The second is to regard the expert assessments as observations and use a Bayesian model to combine them. An even simpler and older method is that of paired comparisons (see the discussion in [27]).

*Pooling.* Key issues of consideration with pooling schemes concern choosing which properties to preserve as we switch between the individual experts and the aggregated pooled expert. For example, assessments which are statistically independent for individuals are not necessarily independent for the pooled expert and updating the pooled expert through Bayes' theorem does not necessarily provide the same distribution as aggregating all the updated individual distributions (see, e.g., [56] or [45]).

We refer to Cooke [27] for a discussion of the different types of pooling, but note that he argued strongly for a linear pooling of expert distributions. Each expert is assigned a weight which is used to form a weighted combination of expert distributions. The weight should not be interpreted as a probability, as one cannot associate the experts with a collection of exclusive and exhaustive events. The choice of weight is difficult to justify. While a common pragmatic approach is to use equal weightings of experts (see the description of NUREG 1150 given in [80]), Cooke has argued that performance-based weighting is more effective and better meets important underlying principles including empirical control. For more details see [10, 27], and for a moment-based approach see [156].

*Bayesian combination.* The main difference in philosophy between pooling and Bayesian combination is that the latter treats expert assessments as if they are observations. Hence there has to be a specification of the likelihood function of the expert data. This feature, which raises serious problems for the analyst, is most clearly visible in the multivariate normal model used by Mosleh and Apostolakis [105]. Here the expert assessment is modeled as being equal to the "true" value of the parameter of interest, plus a normally distributed error, which is considered to be independent of the true value. In principle, the analyst then has to specify the multivariate distribution of expert errors: means (which can be interpreted as expert biases), variances (which can be interpreted as degree of certainty) and covariances (which reflect the degree of correlation of the group of experts). While these are all quantities of some interest in assessing expert opinions, it is not clear on what basis the analyst can assess them without being in a superior position of expertise to that of the experts.

There are now many other Bayesian methods available, especially techniques that incorporate Bayesian networks, which have essentially the same requirement that the analyst develop a likelihood function for expert data. See [24] for a discussion of a variety of such models. The difficulty of structuring such a model depends of course on the details of the model and the context in which it is used. For example, see [129], which takes assessments of numbers of failures to assess parameters of a nonhomogeneous Baysson process, and [57], which discusses the possible advantages of a Bayes linear framework.

4.3.4 *Transformation to parameters and families of distributions.* Both theoreticians and practitioners can easily forget that many of our favorite model parameters, such as failure rate, are not actually observable quantities at all, but are simply parameters of a model that we want to use to make predictions about the future. It has been strongly argued on foundational grounds (e.g., the discussion in Chapter 2 of [10]) that we can only ask for probability assessments on observable quantities. Hence there is a need to infer from those assessments which probability distributions on model parameters are consistent. This approach was developed by Cooke [28]; more algorithms and underlying theory are given in [89].

Taking a more standard Bayesian perspective, Percy [118] discussed the indirect assessment of hyperprior parameters through the direct assessment of quantiles of observables whose distribution is a prior predictive of the unknown Bayesian model. Gutierrez-Pulido, Aguirre-Torres and Christen [59] took a similar line, considering both moments and quantiles of the time to failure for a system as sources of information from which prior distributions can be fitted. Such methods could also be applicable to other Bayesian contexts where prior distributions on lifetime distribution parameters are to be assessed, for example, in Bayesian accelerated or proportional hazards life modeling [22].

In the absence of an assumed class of conditional models it becomes much more difficult to assess a family of conditional distributions: in the context of decision support, it is necessary to consider families of distributions indexed by the decision variables. When the decision space is small and discrete, repeated elicitation can be used, but in the case of a continuous family this becomes more difficult. We are not aware of much work in this area, but it is worth mentioning work by Cooke and Jager, who expressed the probability of an event in terms of system parameters in a Taylor series [26]. A study by Willems et al. [154] used graphical methods to elicit conditional quantiles from experts. In the context of human reliability analysis, proportional hazards type models have been used in which the parameters are assessed purely through various judgemental techniques, such as paired comparison or multicriteria decision analysis. See [42] in particular and the discussion in [10] for other examples.

4.3.5 *The combination of expert and historical data.* Heritage data for historical systems provide insights into the observed reliability of related systems. See, for example, Figure 3, which highlights the selection of historical data to inform the base reliability of the new system.

Historical data may be obtained from generic data bases or company-specific event data bases. Generic reliability data are usually based on operating data drawn from a variety of sources and mixed together. Many generic databases exist; usually they are sector-specific. To adapt such generic information to a system-specific setting, reliability data bases have traditionally used environmental loading factors, such as the *Military Handbook* 217 [37] (hereafter Mil-Hdbk-217).

Mil-Hdbk-217 expresses failure rates for components using so-called "$\pi$" factors, which are multiplication factors that depend on environmental or usage factors. To determine the appropriate failure rate, the analyst simply has to find the correct component description, identify the appropriate environmental or usage factors and then multiply the base failure rate by the $\pi$ factors given in the table. These numbers are given to a very high degree of accuracy and are an attempt to represent the dependence of reliability on at least some parameters. Unfortunately, because they do not represent the dependence on all the parameters, the accuracy given is misleading. By contrast, the IEEE-500 data base and others based on similar principles, such as OREDA and EIREDA, specify much about the system and its operating conditions, but explicitly present the remaining variability in the failure rates. Fragola [50] called these resources "third generation databases."

Many reliability "predictions" made in practical applications seem to be based on an adaptation of generic data through expert opinion, rather than from a (possibly Bayesian-based) fusion of the two forms of data. For example, in practice it is common to adjust generic data to make it system-specific—typically through the use of failure rate multiplication factors—but the methods employed are not generally supported by clear and transparent expert judgement protocols or models.

A nice example of failure rate adjustment is given by Fragola and McFadden's study of failure rates for space station units [51], where experts gather and combine different generic data estimates. While no clear statistical model is used to justify this, it is worth noting that the outputs of the process are *ranges* of failure rates. The third generation databases described above all provide ranges of failure rates—often described using a log-normal distribution on the failure rate parameter. About the point estimates given to great precision in Mil-Hdbk-217 data, Fragola [50] wrote, "...failure rates came to be looked upon as fixed measures of specific equipment, not measures of a spectrum of equipment types." However, he suggested also that Mil-Hdbk-217 data are perfectly usable, as long as they are used in conjunction with uncertainty bands to capture this extra variation.

One of the underlying reasons for the overstated accuracy of Mil-Hdbk-217 is that it reflects a large amount of testing; this is worth reflecting on further, because it has more general implications for the relationship of old to new system data, and for the way data changes through the systems engineering process. Old data will often be a poor representation of prior information because they do not take into account the changes made to the system, usage and environment. Although the usual asymptotic convergence properties hold when updating with the new system data, this is of little practical significance because the amount of new system experience needed for convergence is not available. Hence the speed at which the posteriors will adjust to the "correct" failure rate will be affected by the degree of certainty we had built up for the previous system: The more data we had for the previous system, the more slowly the posterior will converge to the correct new failure rate. A more appropriate way to model the new system reliability is to try to model the change expected to the old system reliability, and this is something that can only be assessed through expert judgement. Often the uncertainty about the effect of the change will dominate the information from the old prior.

The REMM model [148] explicitly attempts to model such effects by considering failure modes in the new and existing systems and subjective assessments about the way the design changes will affect them. A higher level approach using Bayes linear methods (based on moments rather than probabilities) was discussed in [57], where expert assessment of the change to MTBF (mean time before failure) is proposed.

Using historical data from company-specific data bases can give rise to similar issues already mentioned for generic data. Although data for the previous systems manufactured by the company have the potential to relate operating experience directly to earlier design decisions, hence supporting interpretation and selection of base events input to the reliability model for the new system, there are also industry-specific challenges, for example, censoring at the expiry of the warranty period for consumer products [33].

The common problem with using old system data, whatever their source, can be expressed succinctly using the notation for system reliability used earlier,

$$r = r(d, p, u, m, c).$$

Suppose the old system data correspond to slightly different design, production, usage and maintenance patterns. Then the "old" reliability will be

$$r = r(d_o, p_o, u_o, m_o, c_o).$$

The uncertainty ranges given in the third generation data bases may be seen as an attempt to represent credible ranges by changing these parameters within a given envelope. It is finally worth remarking that it is essential for the future utility of these data bases that they maintain the ranges inherent in each equipment class. Hence it would be wrong to start updating the data bases with system-specific data using a straightforward application of Bayes' theorem, as this will reduce the variance artificially.

## 4.4 Revised Quantification with System-Specific Data

Because data are realized from tasks implemented in design, development and manufacture, the initial reliability estimate can be revised as captured in Figures 3–6. We discuss the two options for revision of estimates: Bayesian updating and reelicitation.

As noted above, badly calibrated prior distributions and poorly specified stochastic models compromise Bayesian inference. The quality of inference obtained through Bayesian updating is contingent on both the prior distribution to capture epistemic uncertainty and the choice of model to capture the aleatory uncertainty. The latter of these acts like a lens in which the data are viewed, so even if a meaningful prior distribution is elicited, the posterior distribution may be misleading because this lens may filter out observations that would sensibly inform the inference if the choice of model were different. As such, differences in reelicited prior distributions compared with posterior distributions may be due to the filtering rather than incoherency expert(s) or may be due to a mixture of both.

The systems engineering process is longitudinal and hence offers the opportunity, not only to update prior distributions through Bayes' theorem, but also to reelicit from a common set of experts. This offers the opportunity to validate the choice of model, as well as to assess the calibration of the prior distributions. Furthermore, a learning environment can be created by appropriately feeding results back to the experts and supporting them in improving their ability to specify uncertainty in terms of probability.

As discussed earlier, the quality of subjective probabilities from experts depends on both the elicitation methods and the experts' experience. If an expert lacks experience, prior distributions will be uninformative or misleading, regardless of the elicitation method. Equally, poorly designed elicitation processes may degrade the quality of information provided from experts. Fischhoff [49] proposed the following four necessary conditions to support improving judgement skills: (1) Abundant practice with a set of reasonably homogeneous tasks; (2) clear-cut criterion events for outcome feedback; (3) task-specific reinforcement; (4) explicit admission of the need for learning. There is extensive evidence that these criteria are often not achieved in practice [48, 157].

Feedback is crucial for calibrating the expert and should be event-specific [48, 49, 157]. In other words,

the feedback must be given with respect to assigning probabilities to particular events and not to the ability of the expert to assign probabilities to any situation. To increase the effectiveness of feedback in terms of learning, conditions that influence the event should recur as often as possible [49, 74]. Therefore the factors on which the measure is conditioned should be as few and as general as possible.

## 5. DISCUSSION AND REFLECTIONS ON FUTURE DIRECTIONS

We have attempted to give an overview of expert judgement applications within the field of reliability assessment during systems design. In doing so, a number of key points have arisen which we now revisit to summarize and discuss further.

We have suggested that the whole systems engineering design process is akin to a control problem. The control feedback loops are, however, driven largely not through revisiting decisions in the light of newly acquired system data, but through the use of expert judgements which assess the likely outcome of different system design decisions. Of course, in the wider context of new generations of systems, there are also feedback loops through the use of relevant data that reduce uncertainty not only on the system's own physical and engineering properties, but also on the manner of user interaction. It is also worth mentioning that requirements are frequently revised in light of experience with previous generations of systems, and there is surely a role for statisticians in influencing the setting of such targets. In fact, with the increasing use of sensors that are able to record all sorts of aspects of system performance, environment and use, the opportunities for statistical modeling of these aspects are greater than ever.

Given that systems engineering stresses the importance of making trade-off decisions, we remarked that the reliability information required to support such decisions is—expressed abstractly—the dependence of the reliability metric

$$r = r(d, p, u, m, c)$$

in terms of $d, p, u, m$ and $c$, the choices made for design, production, usage, maintenance and changes. While defining such a function precisely would be impractical, we feel that at least provides a conceptual model for the direction statisticians should be taking. Reliability optimization models, reliability growth models and other such models are all techniques used to provide partial approximations to this function.

The fact that some decisions are made later in the design process means that models are sometimes used in ways that are uncomfortable to statisticians and mathematicians: For example, the use of a constant failure rate lifetime model (exponential distribution) for computations early in the process and the later use of increasing failure rate models for the same system to help determine maintenance intervals may seem contradictory, but if the first model was applied with the knowledge that the maintenance intervals would be fixed post hoc to ensure that the failure rate is roughly constant, then there is no problem. This is a small illustration of the way the flexibility endowed by future decisions can ensure appropriateness of modeling tools post hoc.

Many of the elicitation techniques applied within engineering design have crossed over from the probabilistic risk analysis area. Despite the many similarities with this area, there are a few key differences. One is that many uncertainties will be affected in some way by future design decisions, so an understanding of dependence on design parameters is critical. Another critical aspect is that the design process is one of learning for the engineers. Hence the designers' insights change throughout the process and there is thus a need for problem and model structuring techniques to be applied: the qualitative structuring of statistical models has to be tied closely to the design development process.

Two reliability modeling frameworks have been described that try to extend the scope of reliability techniques from "small world" problems to provide guidance over a wider range of problems. We noted also, however, that a holistic "whole life modeling" approach would need to be attractive to the different stakeholders associated to the system and that there is a need to provide a rational consensus across these parties about the uncertainties faced. Techniques developed in the PRA setting may be adaptable to this situation, and integration with usual systems engineering approaches appears to be natural.

This brings us to some observations on the foundational aspects of reliability modeling in this area, because on the one hand, the methods are largely subjective, but on the other hand, there is a recognition of the limitations of Bayesian techniques. One such limitation is the need to establish rational consensus (as noted above) to break down the sometimes adversarial relationship encountered between stakeholders. Another is that the nature of learning in engineering design is that new modeling needs are continually emerging and model structures—with corresponding likelihood functions—need to be adjusted to match.

Engineering design is an area of great interest for statisticians, but involvement in this area requires some changes in mindset. Reliability is one of the many requirements that the designer is trying to juggle. Hence supporting the design process has to be done by giving insights into what is feasible. All decisions can be modulated later in the process as long as there is a feasible set of solutions: failure of the design process occurs when decisions taken earlier imply that the set of feasible solutions is empty. Historically, many reliability techniques have been applied too little or too late in the design process to inform it properly and some practitioners, such as O'Connor [112], have been critical of statistical reliability work, seeing it as a numbers game, but the increased use of expert judgement combined with more rapid information distribution through information technology systems gives real opportunities to "raise the game" as far as the impact of reliability is concerned. We have identified a whole set of problems in which there is scope for statisticians and operations researchers to play a role in developing new elicitation methods and modeling tools within the systems engineering design process. Surely, as we get more deeply involved in such areas, the insights into uncertainty elicitation will provide benefits for other application areas too. Although we would not go as far as Lindley's colleague in claiming "there are no problems left in statistics except the assessment of probability" [95], it is undeniably the case that expert judgement methods dramatically increase the scope for statistical work in engineering design problems and that work in that area provides us with a new range of contexts within which new elicitation methods can be developed.

## ACKNOWLEDGMENTS

The authors would like to thank numerous colleagues for their suggestions and advice. They would particularly like to thank the engineers involved in the REMM project and J. Fragola for conversations which have helped form an understanding of the nature of uncertainty in the engineering design process.

## REFERENCES

[1] AMARI, S., DILL, G. and HOWALD, E. (2003). A new approach to solve dynamic fault trees. In *Annual Reliability and Maintainability Symposium* 374–379. IEEE Press, Piscataway, NJ.

[2] AMYOT, D., LOGRIPPO, L. and WEISS, M. (2005). Generation of test purposes from Use Case Maps. *Computer Networks* **49** 643–660.

[3] ANDERSON-COOK, C. M., PATTERSON, A. and HOERL, R. (2005). A structured problem-solving course for graduate students: Exposing students to six sigma as part of their university training. *Quality and Reliability Engineering International* **21** 249–256.

[4] ANDREWS, J. D. and MOSS, T. R. (2002). *Reliability and Risk Assessment*, 2nd ed. Professional Engineering Publishing, London.

[5] ANDREWS, J. D. and RIDLEY, L. M. (2002). Application of the cause-consequence diagram method to static systems. *Reliability Engineering and System Safety* **75** 47–58.

[6] ARTHUR, J. D. and GRONER, M. K. (2005). An operational model for structuring the requirements generation process. *Requirements Engineering* **10** 45–62.

[7] AYYUB, B. M. (2001). *Elicitation of Expert Opinions for Uncertainty and Risks.* CRC Press, Boca Raton, FL.

[8] BAGDONAVICIUS, V., BIKELIS, A. and KAZAKEVICIUS, V. (2005). Nonparametric confidence intervals for the number of spares required for a renewable system. *Comm. Statist. Theory Methods* **34** 1203–1212. MR2189427

[9] BARENDREGT, A., VAN NOORTWIJK, J. M., VAN DER DOEF, M. and HOLTERMAN, S. R. (2005). Determining the time available for evacuation of a dike-ring area by expert judgement. In *Proc. Ninth International Symposium on Stochastic Hydraulics (ISSH)* (J. K. Vrijling, E. Ruijgh, B. Stalenberg, P. H. A. J. M. van Gelder, M. Verlaan, A. Zijderveld and P. Waarts, eds.). International Association of Hydraulic Engineering and Research (IAHR), Nijmegen, Netherlands.

[10] BEDFORD, T. and COOKE, R. (2001). *Probabilistic Risk Analysis: Foundations and Methods.* Cambridge Univ. Press. MR1825639

[11] BEDFORD, T. and QUIGLEY, J. (2004). Risk reduction prioritisation using decision analysis. *Risk, Decision and Policy* **9** 223–236.

[12] BENNETT, T. R., BOOKER, J. M., KELLER-MCNULTY, S. and SINGPURWALLA, N. D. (2003). Testing the untestable: Reliability in the 21st century. *IEEE Transactions on Reliability* **52** 118–124.

[13] BIERBAUM, R. L., BROWN, T. D. and KERSCHEN T. J. (2002). Model-based reliability analysis. *IEEE Transactions on Reliability* **51** 133–140.

[14] BLANCHARD, B. S. (2004). *System Engineering Management*, 3rd ed. Wiley, Hoboken, NJ.

[15] BLANCHARD, B. S. and FABRYCKY, W. J. (1998). *Systems Engineering and Analysis*, 3rd ed. Prentice Hall, Upper Saddle River, NJ.

[16] BLISCHKE, W. R. and MURTHY, D. N. P. (2000). *Reliability: Modeling, Prediction and Optimization*. Wiley, New York.

[17] BOOKER, J. D. and MCNAMARA, L. A. (2002). Expertise and expert judgment in reliability characterization: A rigorous approach to eliciting, documenting and analyzing expert knowledge. In *Engineering Design and Reliability Handbook* (E. Nikolaidis, D. Ghiocel and S. Singhal, eds.) Chapter 13. CRC Press, Boca Raton, FL. MR2186556

[18] BOUDALI, H. and DUGAN, J. B. (2005). A discrete-time Bayesian network reliability modeling and analysis framework. *Reliability Engineering and System Safety* **87** 337–349.

[19] BOUISSOU, M., MARTIN, F. and OURGHANLIAN, A. (1999). Assessment of a safety-critical system including software: A Bayesian belief network for evidence sources. In *Annual Reliability and Maintainability Symposium* 142–150. IEEE Press, Piscataway, NJ.

[20] BOWLES, J. B. (2000). Designing for failure: Managing the failure response through analysis. In *Failure Prevention through Education: Getting at the Root Cause* 62–69. ASM International, Materials Park, OH.

[21] BOWLES, J. B. (2003). An assessment of RPN prioritization in a failure modes effects and criticality analysis. In *Annual Reliability and Maintainability Symposium* 380–386. IEEE Press, Piscataway, NJ.

[22] BUNEA, C. and MAZZUCHI, T. A. (2005). Bayesian accelerated life testing under competing failure modes. In *Annual Reliability and Maintainability Symposium* 152–157. IEEE Press, Piscataway, NJ.

[23] CLAUSING, D. (1994). *Total Quality Development: A Step-by-Step Guide to World Class Concurrent Engineering*. ASME Press, New York.

[24] CLEMEN, R. T. and WINKLER, R. L. (1999). Combining probability distributions from experts in risk analysis. *Risk Analysis* **19** 187–203.

[25] CONDRA, L. (1993). *Reliability Improvement with Design of Experiments*. Dekker, New York.

[26] COOKE, R. and JAGER, E. (1998). Probabilistic model for the failure frequency of underground gas pipelines. *Risk Analysis* **18** 511–527.

[27] COOKE, R. M. (1991). *Experts in Uncertainty*. Oxford Univ. Press. MR1136548

[28] COOKE, R. M. (1994). Parameter fitting for uncertain models: Modelling uncertainty in small models. *Reliability Engineering and System Safety* **44** 89–102.

[29] COOKE, R. M. (2003). Book review of *Elicitation of Expert Opinions for Uncertainty and Risks*, by B. M. Ayyub. *Fuzzy Sets and Systems* **133** 267–268.

[30] COOKE, R. M. and SLIJKHUIS, K. A. (2003). Expert judgment in the uncertainty analysis of dike ring failure frequency. In *Case Studies in Reliability and Maintenance* (W. R. Blischke and D. N. P. Murthy, eds.) 331–350. Wiley, Hoboken, NJ. MR1959779

[31] COWING, M. M., PATÉ-CORNELL, M. E. and GLYNN, P. W. (2004). Dynamic modeling of the tradeoff between productivity and safety in critical engineering systems. *Reliability Engineering and System Safety* **86** 269–284.

[32] CYSNEIROS, L. M. and YU, E. (2003). Requirements engineering for large-scale multi-agent systems. *Software Engineering for Large-Scale Multi-Agent Systems. Lecture Notes in Comput. Sci.* **2603** 39–56. Springer, New York.

[33] DAVIS, T. P. (2003). Reliability improvement in automotive engineering. In *Global Vehicle Reliability: Prediction and Optimization Techniques* (J. E. Strutt and P. L. Hall, eds.) Chapter 4. Professional Engineering Publishing, London.

[34] DAWSON, R., HALL, J. and DAVIS, J. (2004). Performance-based management of flood defence systems. *Proc. ICE, Water Management* **157** 35–44.

[35] DE LANDTSHEER, R., LETIER, E. and VAN LAMSWEERDE, A. (2004). Deriving tabular event-based specifications from goal-oriented requirements models. *Requirements Engineering* **9** 104–120.

[36] DE WIT, S. and AUGENBROE, G. (2002). Analysis of uncertainty in building design evaluations and its implications. *Energy and Buildings* **34** 951–958.

[37] DEPARTMENT OF DEFENSE (1991). Military handbook: Reliability prediction of electronic equipment.

[38] DOGGETT, A. M. (2004). A statistical comparison of three root cause analysis tools. *J. Industrial Technology* **20** (2), 9 pp.

[39] DONZELLI, P. (2004). A goal-driven and agent-based requirements engineering framework. *Requirements Engineering* **9** 16–39.

[40] EBRAHIMI, N. B. (2003). Indirect assessment of system reliability. *IEEE Transactions on Reliability* **52** 58–62.

[41] EDEN, C. (2004). Analyzing cognitive maps to help structure issues or problems. *European J. Oper. Res.* **159** 673–686.

[42] EMBREY, D. E., HUMPHREYS, P. C., ROSA, E. A., KIRWAN, B. and REA, K. (1984). SLIM-MAUD: An approach to assessing human error probabilities using structured expert judgement. Technical report NUREG/CR-3518, U.S. Nuclear Regulatory Commission.

[43] EPSTEIN, S. and RAUZY, A. (2005). Can we trust PRA? *Reliability Engineering and System Safety* **88** 195–205.

[44] ERKANLI, A., MAZZUCHI, T. A. and SOYER, R. (1998). Bayesian computations for a class of reliability growth models. *Technometrics* **40** 14–23. MR1625623

[45] FARIA, A. E. and SMITH, J. Q. (1997). Conditionally externally Bayesian pooling operators in chain graphs. *Ann. Statist.* **25** 1740–1761. MR1463573

[46] FENTON, N., LITTLEWOOD, B., NEIL, M., STRIGINI, L., SUTCLIFFE, A. and WRIGHT, D. (1998). Assessing dependability of safety critical systems us-

ing diverse evidence. *IEE Proceedings Software* **145** 35–39.

[47] FENTON, N. and NEIL, M. (2001). Making decisions: Using Bayesian nets and MCDA. *Knowledge-Based Systems* **14** 307–325.

[48] FERRELL, W. (1994). Discrete subjective probabilities and decision analysis: Elicitation, calibration and combination. In *Subjective Probability* (G. Wright and P. Ayton, eds.) 411–451. Wiley, New York. MR1307663

[49] FISCHHOFF, B. (1989). Eliciting knowledge for analytical representation. *IEEE Transactions on Systems, Man and Cybernetics* **19** 448–461.

[50] FRAGOLA, J. R. (1996). Reliability and risk analysis data base development: An historical perspective. *Reliability Engineering and System Safety* **51** 125–136.

[51] FRAGOLA, J. R. and MCFADDEN, R. H. (1995). Synthesis of failure rates for space station external orbital replaceable units. *Reliability Engineering and System Safety* **49** 237–253.

[52] FRANK, M. V. (2005). View through the door of the SOFIA project. *IEEE Transactions on Reliability* **54** 181–188.

[53] GARTHWAITE, P. H., KADANE, J. B. and O'HAGAN, A. (2005). Statistical methods for eliciting probability distributions. *J. Amer. Statist. Assoc.* **100** 680–700. MR2170464

[54] GARVIN, T. (2001). Analytical paradigms: The epistemological distances between scientists, policy makers and the public. *Risk Analysis* **21** 443–455.

[55] GEBRAEEL, N. Z., LAWLEY, M. A., LI, R. and RYAN, J. K. (2005). Residual-life distributions from component degradation signals: A Bayesian approach. *IIE Transactions* **37** 543–557.

[56] GENEST, C., MCCONWAY, K. J. and SCHERVISH, M. J. (1986). Characterization of externally Bayesian pooling operators. *Ann. Statist.* **14** 487–501. MR0840510

[57] GOLDSTEIN, M. and BEDFORD, T. (2005). The Bayes linear approach to inference and decision making for a reliability programme. Technical Report 2005/14, Strathclyde Univ.

[58] GUIDA, M. and PULCINI, G. (2005). Bayesian reliability assessment of repairable systems during multistage development programs. *IIE Transactions* **37** 1071–1081.

[59] GUTIERREZ-PULIDO, H., AGUIRRE-TORRES, V. and CHRISTEN, J. A. (2005). A practical method for obtaining prior distributions in reliability. *IEEE Transactions on Reliability* **54** 262–269.

[60] HAHN, G. J. (2005). Six sigma: 20 key lessons learned. *Quality and Reliability Engineering International* **21** 225–233.

[61] HALL, J. W., MEADOWCROFT, I. C., LEE, E. M. and VAN GELDER, P. H. A. J. M. (2002). Stochastic simulation of episodic soft coastal cliff recession. *Coastal Engineering* **46** 159–174.

[62] HODGE, R., EVANS, M., MARSHALL, J., QUIGLEY, J. and WALLS, L. (2001). Eliciting engineering knowledge about reliability during design-lessons learnt from implementation. *Quality and Reliability Engineering International* **17** 169–179.

[63] HOWICK, S. (2003). Using system dynamics to analyse disruption and delay in complex projects for litigation: Can the modelling purposes be met? *J. Oper. Res. Soc.* **54** 222–229.

[64] HOWICK, S. (2005). Using system dynamics models with litigation audiences. *European J. Oper. Res.* **162** 239–250.

[65] HOWICK, S. and EDEN, C. (2001). The impact of disruption and delay in large projects: Going for incentives? *J. Oper. Res. Soc.* **52** 26–34.

[66] HOWICK, S. and EDEN, C. (2004). On the nature of discontinuities in system dynamics modelling of disrupted projects. *J. Oper. Res. Soc.* **55** 598–605.

[67] HSU, L. F. (2005). Group testing with a goal in estimating the number of defects under imperfect environmental stress screen levels. *Comm. Statist. Theory Methods* **34** 1363–1377. MR2162096

[68] HUANG, Y.-S. and ZHUO, Y.-F. (2004). Estimation of future breakdowns to determine optimal warranty policies for products with deterioration. *Reliability Engineering and System Safety* **84** 163–168.

[69] IEC 60300 SERIES. International Electrotechnical Commission, Geneva.

[70] INTERNATIONAL ELECTROTECHNICAL COMMISSION (1991). International electrotechnical vocabulary.

[71] JAMES, I., MARSHALL, J., EVANS, M. and NEWMAN, B. (2004). Reliability metrics and the REMM model. In *Annual Reliability and Maintainability Symposium* 474–479. IEEE Press, Piscataway, NJ.

[72] JENKINSON, D. (2005). The elicitation of probabilities—a review of the statistical literature. BEEP working paper, Univ. Sheffield.

[73] JOHNSON, V. E., MOOSMAN, A. and COTTER, P. (2005). A hierarchical model for estimating the early reliability of complex systems. *IEEE Transactions on Reliability* **54** 224–231.

[74] KADANE, J. B. and WOLFSON, L. J. (1998). Experiences in elicitation. *The Statistician* **47** 3–19.

[75] KAHNEMAN, D., SLOVIC, P. and TVERSKY, A., eds. (1982). *Judgment under Uncertainty: Heuristics and Biases.* Cambridge Univ. Press.

[76] KALLEN, M. J. and COOKE, R. M. (2002). Expert aggregation with dependence. In *Probabilistic Safety Assessment and Management* (E. J. Bonano, A. L. Camp, M. J. Majors and R. A. Thompson, eds.) 1287–1294. North-Holland, Amsterdam.

[77] KAMINSKIY, M. P. and KRIVTSOV, V. V. (2005). A simple procedure for Bayesian estimation of the Weibull distribution. *IEEE Transactions on Reliability* **54** 612–616.

[78] KANJI, G. K. and ASHER, M. (1996). *100 Methods for Total Quality Management.* Sage, London.

[79] KEARNEY, M., MARSHALL, J. and NEWMAN, B. (2003). Comparison of reliability enhancement tests for electronic equipment. In *Annual Reliability and*

*Maintainability Symposium* 435–440. IEEE Press, Piscataway, NJ.

[80] KEENEY, R. L. and VON WINTERFELDT, D. (1991). Eliciting probabilities from experts in complex technical problems. *IEEE Transactions on Engineering Management* **38** 191–201.

[81] KELLER, W. W. and MODARRES, M. (2005). A historical overview of probabilistic risk assessment development and its use in the nuclear power industry: A tribute to the late Professor Norman Carl Rasmussen. *Reliability Engineering and System Safety* **89** 271–285.

[82] KELLER-MCNULTY, S. (2004). Reliability for the 21st century. In *Mathematical Methods in Reliability* 1–4.

[83] KERSCHER, W. J., BOOKER, J. M., BEMENT, T. R. and MEYER M. A. (1998). Characterizing reliability in a product/process design-assurance program. In *Annual Reliability and Maintainability Symposium* 105–112. IEEE Press, Piscataway, NJ.

[84] KERSCHER, W. J., BOOKER, J. M., MEYER, M. A. and SMITH, R. E. (2003). PREDICT: A case study, using fuzzy logic. In *Annual Reliability and Maintainability Symposium* 188–195. IEEE Press, Piscataway, NJ.

[85] KLETZ, T. A. (1997). Hazop—past and future. *Reliability Engineering and System Safety* **55** 263–266.

[86] KLEYNER, A. and SANDBORN, P. (2005). A warranty forecasting model based on piecewise statistical distributions and stochastic simulation. *Reliability Engineering and System Safety* **88** 207–214.

[87] KOPNOV, V. A. (1999). Optimal degradation processes control by two-level policies. *Reliability Engineering and System Safety* **66** 1–11.

[88] KORVING, H. and VAN NOORTWIJK, J. M. (2006). Bayesian updating of a prediction model for sewer degradation. In *2nd International IWA Conference on Sewer Operation and Maintenance*. Vienna.

[89] KRAAN, B. and BEDFORD, T. (2005). Probabilistic inversion of expert judgments in the quantification of model uncertainty. *Management Science* **51** 995–1006. MR2230697

[90] KUNTTU, S. and KORTELAINEN, H. (2004). Supporting maintenance decisions with expert and event data. In *Annual Reliability and Maintainability Symposium* 593–599. IEEE Press, Piscataway, NJ.

[91] KUO, W. and PRASAD, V. R. (2000). An annotated overview of system-reliability optimization. *IEEE Transactions on Reliability* **49** 176–187.

[92] LAWLESS, J. (2000). Statistics in reliability. *J. Amer. Statist. Assoc.* **95** 989–992.

[93] LEE, B. H. (2001). Using FMEA models and ontologies to build diagnostic models. *Artificial Intelligence for Engineering Design, Analysis and Manufacturing* **15** 281–293.

[94] LEISHMAN, D. and MCNAMARA, L. (2002). Interlopers, translators, scribes and seers: Anthropology, knowledge representation and Bayesian statistics for predictive modeling in multidisciplinary science and engineering projects. In *Proc. Conference on Visual Representations and Interpretations 2002*. Liverpool.

[95] LINDLEY, D. V. (2000). The philosophy of statistics (with discussion). *The Statistician* **49** 293–337.

[96] LU, W. Y., WANG, W. and CHRISTER A. H. (2005). The delay time modelling of preventive maintenance of plant based on subjective PM data and actual failure records. In *Proc. 4th International Conference on Quality and Reliability* 805–812.

[97] LUTZ, R. R. and MIKULSKI, I. C. (2003). Operational anomalies as a cause of safety-critical requirements evolution. *J. Systems and Software* **65** 155–161.

[98] MAIDEN, N., GIZIKIS, A. and ROBERTSON, S. (2004). Provoking creativity: Imagine what your requirements could be like. *IEEE Software* **21** (5) 68–75.

[99] MARQUEZ, A. C., HEGUEDAS, A. S. and IUNG, B. (2005). Monte Carlo-based assessment of system availability. A case study for cogeneration plants. *Reliability Engineering and System Safety* **88** 273–289.

[100] MARSHALL, J. and NEWMAN, R. (1998). Reliability enhancement methodology and modeling for electronic equipment—the REMM project. In *Proc. ERA Avionics Conference and Exhibition* 4.2.1–4.2.13.

[101] MATHEWSON, A., O'SULLIVAN, P., CONCANNON, A., FOLEY, S., MINEHANE, S., DUANE, R. and PALSER, K. (1999). Modelling and simulation of reliability for design. *Microelectronic Engineering* **49** 95–117.

[102] MEEKER, W. Q. and ESCOBAR, L. A. (2004). Reliability: The other dimension of quality. *Qual. Technol. Quant. Manag.* **1** 1–25. MR2190374

[103] MERKHOFER M. W. (1987). Quantifying judgmental uncertainty: Methodology, experiences and insights. *IEEE Transactions on Systems, Man and Cybernetics* **17** 741–752.

[104] MEYER, M. A. and BOOKER, J. M. (2001). *Eliciting and Analyzing Expert Judgment: A Practical Guide*, 2nd ed. SIAM, Philadelphia. MR1823794

[105] MOSLEH, A. and APOSTOLAKIS, G. (1986). The assessment of probability distributions form expert opinions with an application to seismic fragility curves. *Risk Analysis* **6** 447–461.

[106] MURTHY, D. N. P., SOLEM, O. and ROREN, T. (2004). Product warranty logistics: Issues and challenges. *European J. Oper. Res.* **156** 110–126.

[107] NEIL, M., LITTLEWOOD, B. and FENTON, N. (1996). Applying Bayesian belief networks to system dependability assessment. In *Proc. Safety Critical Systems Club Symposium* 84–95. Springer, Berlin.

[108] NEILL, C. J. and LAPLANTE, P. A. (2003). Requirements engineering: The state of the practice. *IEEE Software* **20** (6) 40–45.

[109] NELSON, W. B. (2005). A bibliography of accelerated test plans. *IEEE Transactions on Reliability* **54** 194–197.

[110] NELSON, W. B. (2005). A bibliography of accelerated test plans. II. *IEEE Transactions on Reliability* **54** 370–373.

[111] O'CONNOR, P. D. T. (1991). *Practical Reliability Engineering*, 3rd ed. Wiley, New York.

[112] O'CONNOR, P. D. T. (2000). Commentary: Reliability—past, present and future. *IEEE Transactions on Reliability* 49 335–341.

[113] O'HAGAN, A. (1998). Eliciting expert beliefs in substantial practical applications. *The Statistician* 47 21–35.

[114] O'HAGAN, A. (2005). Research in elicitation. Technical Report 557/05, Dept. Probability and Statistics, Univ. Sheffield.

[115] O'HAGAN, A. and OAKLEY, J. E. (2004). Probability is perfect, but we can't elicit it perfectly. *Reliability Engineering and System Safety* 85 1–3, 239–248.

[116] PARK, G.-J., LEE, T.-H., LEE, K. H. and HWANG, K. H. (2006). Robust design: An overview. *AIAA Journal* 44 181–191.

[117] PATTERSON, A., BONISSONE, P. and PAVESE, M. (2005). Six sigma applied throughout the lifecycle of an automated decision system. *Quality and Reliability Engineering International* 21 275–292.

[118] PERCY, D. F. (2002). Bayesian enhanced strategic decision making for reliability. *European J. Oper. Res.* 139 133–145. MR1888266

[119] PETKOVA, V. T., YUAN, L., ION, R. A. and SANDER, P. C. (2005). Designing reliability information flows. *Reliability Engineering and System Safety* 88 147–155.

[120] PHILLIPS, L. D. (1982). Requisite decision modeling: A case study. *J. Oper. Res. Soc.* 33 303–311.

[121] PHILLIPS, L. D. (1984). A theory of requisite decision models. *Acta Psychologica* 56 29–48.

[122] PHILLIPS, M. J. (2004). Bayesian estimation from censored data with incomplete information. *Quality and Reliability Engineering International* 20 237–245.

[123] PIDD, M. (2003). *Tools for Thinking: Modelling in Management Science*, 2nd ed. Wiley, Hoboken, NJ.

[124] QUIGLEY, J. and WALLS, L. (1999). Measuring the effectiveness of reliability growth testing. *Quality and Reliability Engineering International* 15 87–93.

[125] QUIGLEY, J. and WALLS, L. (2003). Confidence intervals for reliability-growth models with small sample-sizes. *IEEE Transactions on Reliability* 52 257–262.

[126] RAUSAND, M. and HØYLAND, A. (2004). *System Reliability Theory: Models, Statistical Methods and Applications*, 2nd ed. Wiley, Hoboken, NJ. MR2016162

[127] RENN, O. (2004). The role of stakeholder involvement in risk communication. *Internationale Zeitschrift für Kernenergie* 49 602–612.

[128] RENN, O. and KLINKE, A. (2004). Systemic risks: A new challenge for risk management. *EMBO Reports* 5 S41–S46.

[129] ROSQVIST, T. (2000). Bayesian aggregation of experts' judgements on failure intensity. *Reliability Engineering and System Safety* 70 283–289.

[130] RUAN, J., SHAYA, N., CASSAR, V., PATTINSON, C., ASH-RAFZADEH, A. and PRASAD, P. (2002). The house of quality and an enhanced vehicle attributes design process. In *Proc. Fifth International Conference on Frontiers of Design and Manufacturing* 1 106–113.

[131] SALEH, J. H. and MARAIS, K. (2006). Highlights from the early (and pre-) history of reliability engineering. *Reliability Engineering and System Safety* 91 249–256.

[132] SHISHKO, R. and CHAMBERLAIN, R. G. (1995). *NASA Systems Engineering Handbook*.

[133] SHRADER-FRECHETTE, K. S. (1991). *Risk and Rationality*. Univ. California Press, Berkeley.

[134] SIGURDSSON, J. H., WALLS, L. A. and QUIGLEY, J. L. (2001). Bayesian belief nets for managing expert judgement and modelling reliability. *Quality and Reliability Engineering International* 17 181–190.

[135] SINGPURWALLA, N. D. (1995). Survival in dynamic environments. *Statist. Sci.* 10 86–103.

[136] SLOVIC, P., FINUCANE, M. L., PETERS, E. and MACGREGOR, D. G. (2004). Risk as analysis and risk as feelings: Some thoughts about affect, reason, risk and rationality. *Risk Analysis* 24 311–322.

[137] STALLINGER, F. and GRÜNBACHER, P. (2001). System dynamics modelling and simulation of collaborative requirements engineering. *J. Systems and Software* 59 311–321.

[138] STEPHENS, D. and CROWDER, M. (2004). Bayesian analysis of discrete time warranty data. *Appl. Statist.* 53 195–217. MR2037885

[139] TAN, C. M. and NEO, T.-K. (2002). QFD implementation in a discrete semiconductor industry. In *Annual Reliability and Maintainability Symposium* 484–489. IEEE Press, Piscataway, NJ.

[140] TORRES-TOLEDANO, J. G. and SUCAR, L. E. (1998). Bayesian networks for reliability analysis of complex systems. *Progress in Artificial Intelligence–Iberamia 98. Lecture Notes in Comput. Sci.* 1484 195–206. Springer, New York.

[141] VAN ELST, N., BEDFORD, T., JORISSEN, R. and KLAASSEN, D. (1998). A generic risk model for the closing procedure of moveable water barriers. In *Safety and Reliability, Proc. ESREL '98* (H. A. Sandtorv, ed.) 2 435–442. Balkema, Rotterdam.

[142] VAN NOORTWIJK, J. M., DEKKER, A., COOKE, R. M. and MAZZUCHI, T. A. (1992). Expert judgment in maintenance optimization. *IEEE Transactions on Reliability* 41 427–432.

[143] VAURIO, J. K. (2001). Fault tree analysis of phased mission systems with repairable and non-repairable components. *Reliability Engineering and System Safety* 74 169–180.

[144] VLOK, P. J., COETZEE, J. L., BANJEVIC, D., JARDINE, A. K. S. and MAKIS, V. (2002). Optimal component replacement decisions using vibration monitoring and the proportional-hazards model. *J. Oper. Res. Soc.* 53 193–202.

[145] VROUWENVELDER, A. C. W. M. (2003) Uncertainty analysis for flood defence systems in The Nether-

lands. In *Safety and Reliability: Proc. ESREL 2003* (T. Bedford and P. van Gelder, eds.) **1** 11–18. Balkema, Maastricht, The Netherlands.

[146] WALLS, L. and QUIGLEY, J. (2001). Building prior distributions to support Bayesian reliability growth modelling using expert judgement. *Reliability Engineering and System Safety* **74** 117–128.

[147] WALLS, L., QUIGLEY, J. and KRASICH, M. (2005). Comparison of two models for managing reliability growth during product design. *IMA J. Management Mathematics* **16** 12–22.

[148] WALLS, L., QUIGLEY, J. and MARSHALL, J. (2006). Modeling to support reliability enhancement during product development with applications in the U.K. aerospace industry. *IEEE Transactions on Engineering Management* **53** 263–274.

[149] WANG, W. (1997). Subjective estimation of the delay time distribution in maintenance modelling. *European J. Oper. Res.* **99** 516–529.

[150] WARD, H. and CHRISTER A. H. (2005). Modelling the re-design decision for a warranted product. *Reliability Engineering and System Safety* **88** 181–189.

[151] WATAHIKI, K. and SAEKI, M. (2004). Combining goal-oriented analysis and use case analysis. *IEICE Transactions on Information and Systems* **E87D** 822–830.

[152] WEBER, P. and JOUFFE, L. (2006). Complex system reliability modelling with Dynamic Object Oriented Bayesian Networks (DOOBN). *Reliability Engineering and System Safety* **91** 149–162.

[153] WHITE, D. (1995). Application of systems thinking to risk management: A review of the literature. *Management Decision* **33** (10) 35–45.

[154] WILLEMS, A., JANSSEN, M., VERSTEGEN, C. and BEDFORD, T. (2005). Expert quantification of uncertainties in a risk analysis for an infrastructure project. *J. Risk Research* **8** 3–17.

[155] WILSON, A., MCNAMARA, L. and WILSON, G. (2004). Information integration for complex systems. Technical Report LA-UR-04-6564, Los Alamos National Laboratory.

[156] WISSE, B., BEDFORD, T. and QUIGLEY, J. (2005). Combining expert judgements in the Bayes linear methodology. In *Proc. CEA-JRC Workshop on the Use of Expert Judgement in Decision-Making* (N. Devictor, V. Moulin and R. Bolado-Lavin, eds.). CEC, Aix-en-Provence.

[157] WRIGHT, G. and BOLGER, F., eds. (1992). *Expertise and Decision Support.* Plenum, New York.

[158] YOO, J., CATANIO, J., PAUL, R. and BIEBER, M. (2004). Relationship analysis in requirements engineering. *Requirements Engineering* **9** 238–247.

[159] YU, D. C., NGUYEN, T. C. and HADDAWY, P. (1999). Bayesian network model for reliability assessment of power systems. *IEEE Transactions on Power Systems* **14** 426–432.